# IceCube

**Albrecht Karle**[*], **for the IceCube Collaboration**

[*]*University of Wisconsin-Madison, 1150 University Avenue, Madison, WI 53706*

*Abstract.* IceCube is a 1 km$^3$ neutrino telescope currently under construction at the South Pole. The detector will consist of 5160 optical sensors deployed at depths between 1450 m and 2450 m in clear Antarctic ice evenly distributed over 86 strings. An air shower array covering a surface area of 1 km$^2$ above the in-ice detector will measure cosmic ray air showers in the energy range from 300 TeV to above 1 EeV. The detector is designed to detect neutrinos of all flavors: $\nu_e$, $\nu_\mu$ and $\nu_\tau$. With 59 strings in operation in 2009, construction is 67% complete. Based on data taken to date, the observatory meets its design goals. Selected results will be presented.

## I. INTRODUCTION

IceCube is a large kilometer scale neutrino telescope currently under construction at the South Pole. With the ability to detect neutrinos of all flavors over a wide energy range from about 100 GeV to beyond $10^9$ GeV, IceCube is able to address fundamental questions in both high energy astrophysics and neutrino physics. One of its main goals is the search for sources of high energy astrophysical neutrinos which provide important clues for understanding the origin of high energy cosmic rays.

The interactions of ultra high energy cosmic rays with radiation fields or matter either at the source or in intergalactic space result in a neutrino flux due to the decays of the produced secondary particles such as pions, kaons and muons. The observed cosmic ray flux sets the scale for the neutrino flux and leads to the prediction of event rates requiring kilometer scale detectors, see for example[1]. As primary candidates for cosmic ray accelerators, AGNs and GRBs are thus also the most promising astrophysical point source candidates of high energy neutrinos. Galactic source candidates include supernova remnants, microquasars, and pulsars. Guaranteed sources of neutrinos are the cosmogenic high energy neutrino flux from interactions of cosmic rays with the cosmic microwave background and the galactic neutrino flux resulting from galactic cosmic rays interacting with the interstellar medium. Both fluxes are small and their measurement constitutes a great challenge. Other sources of neutrino radiation include dark matter, in the form of supersymmetric or more exotic particles and remnants from various phase transitions in the early universe.

The relation between the cosmic ray flux and the atmospheric neutrino flux is well understood and is based on the standard model of particle physics. The observed diffuse neutrino flux in underground laboratories agrees with Monte Carlo simulations of the primary cosmic ray flux interacting with the Earth's atmosphere and producing a secondary atmospheric neutrino flux[2].

Although atmospheric neutrinos are the primary background in searching for astrophysical neutrinos,

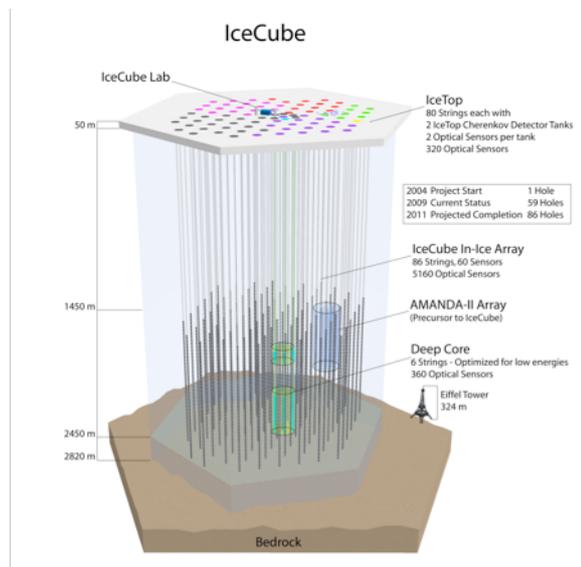

Fig. 1 Schematic view of IceCube. Fifty-nine of 86 strings are in operation since 2009.

they are very useful for two reasons. Atmospheric neutrino physics can be studied up to PeV energies. The measurement of more than 50,000 events per year in an energy range from 500 GeV to 500 TeV will make IceCube a unique instrument to make precise comparisons of atmospheric neutrinos with model predictions. At energies beyond 100 TeV a harder neutrino spectrum may emerge which would be a signature of an extraterrestrial flux. Atmospheric neutrinos also give the opportunity to calibrate the detector. The absence of such a calibration beam at higher energies poses a difficult challenge for detectors at energies targeting the cosmogenic neutrino flux.

## II. DETECTOR AND CONSTRUCTION STATUS

IceCube is designed to detect muons and cascades over a wide energy range. The string spacing was chosen in order to reliably detect and reconstruct muons in the TeV energy range and to precisely calibrate the detector using flashing LEDs and atmospheric muons. The optical properties of the South Pole Ice have been measured with various calibration devices[3] and are used for modeling the detector response to charged particles. Muon reconstruction algorithms[4] allow measuring the direction and energy of tracks from all directions.

In its final configuration, the detector will consist of 86 strings reaching a depth of 2450 m below the surface. There are 60 optical sensors mounted on each string equally spaced between 1450m and 2450m depth with the exception of the six Deep Core strings on which the sensors are more closely spaced between 1760m and 2450m. In addition there will be 320 sensors deployed in 160 IceTop tanks on the surface of the ice directly above the strings. Each sensor consists of a 25cm photomultiplier tube (PMT), connected to a waveform recording data acquisition circuit capable of resolving pulses with nanosecond precision and having a dynamic range of at least 250 photoelectrons per 10ns.

The detector is constructed by drilling holes in the ice, one at a time, using a hot water drill. Drilling is immediately followed by deployment of a detector string into the water-filled hole. The drilling of a hole to a depth of 2450m takes about 30 hours. The subsequent deployment of the string typically takes less than 10 hours. The holes typically freeze back within 1-3 weeks. The time delay between two subsequent drilling cycles and string deployments was in some cases shorter than 50 hours. By the end of February 2009, 59 strings and IceTop stations had been deployed. We refer to this configuration as IC59. Once the strings are completely frozen in the commissioning can start. Approximately 99% of the deployed DOMs have been successfully commissioned. The 40-string detector configuration (IC40) has been in operation from May 2008 to the end of April 2009.

## III. MUONS AND NEUTRINOS

At the depth of IceCube, the event rate from downgoing atmospheric muons is close to 6 orders of magnitude higher than the event rate from atmospheric neutrinos. Fig. 2 shows the observed muon rate (IC22) as a function of the zenith angle[5]. IceCube is effective in detecting downward going muons. A first measurement of the muon energy spectrum is provided in the references[6].

A good angular resolution of the experiment is the basis for the zenith angle distribution and much more so for the search of point sources of neutrinos from galactice sources, AGNs or GRBs. Figure 3 shows the angular resolution of IceCube for several detector configurations based on high quality neutrino event selections as used in the point source search for IC40 [7]. The median angular resolution of IC40 achieved already 0.7°, the design parameter for the full IceCube.

The muon flux serves in many ways also as a calibration tool. One method to verify the angular resolution and absolute pointing of the detector uses the Moon shadow of cosmic rays. The Moon reaches an elevation of about 28° above the horizon at the South Pole. Despite the small altitude of the Moon, the event rate and angular resolution of IceCube are sufficient to measure the cosmic ray shadow of the Moon by mapping the muon rate in the vicinity of the Moon. The parent air showers have an energy of typically 30 TeV, well above the energy where magnetic fields would pose a significant deviation from the direction of the primary par-ticles. Fig. 4 shows a simple declination band with bin size optimized for this analysis. A deficit of ~900 events (~5σ) is observed on a background of ~28000 events in 8 months of data taking. The deficit is in agreement with expectations and confirms the assumed angular resolution and absolute pointing.

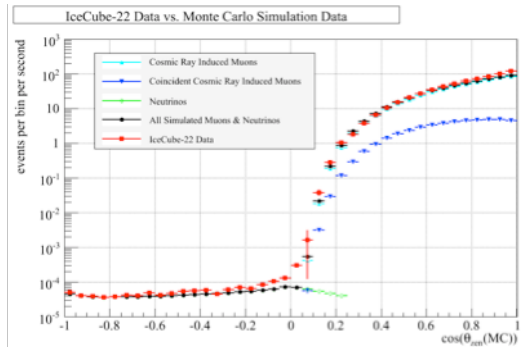

Fig. 2 Muon rate in IceCube as a function of zenith angle [55]. The data agree with the detector simulation which includes atmospheric neutrinos, atmospheric muons, and coincident cosmic ray muons (two muons erroneously reconstructed as a single track.)

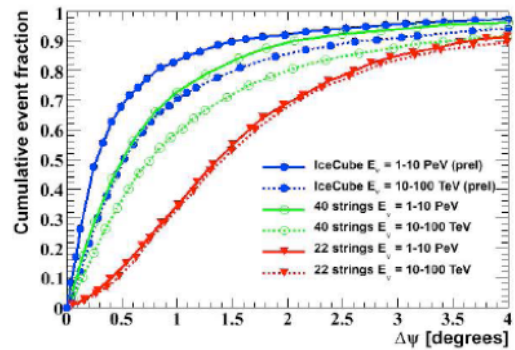

Fig. 3 The angular resolution function of different IceCube configurations is shown for two neutrino energy ranges samples from an $E^{-2}$ energy spectrum.

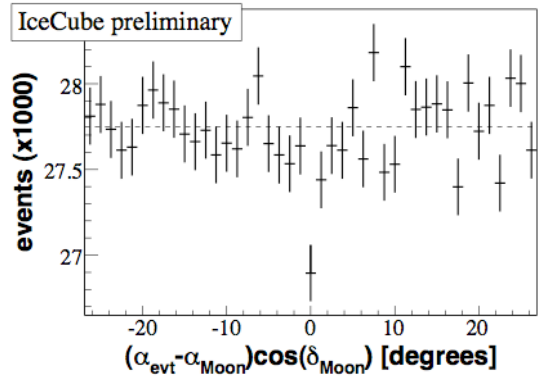

Fig. 4: A 4.2σ deficit of events from direction of Moon in the IceCube 40-string detector confirms pointing accuracy.

The full IceCube will collect of order 50 000 high quality atmospheric neutrinos per year in the TeV energy range. A detailed understanding of the response function of the detector at analysis level is the foundation for any neutrino flux measurement. We use the concept of the neutrino effective area to describe the response function of the detector with respect to neutrino flavor, energy and zenith angle. The neutrino effective area is the equivalent area for which all neutrinos of a given neutrino flux impinging on the Earth would be observed. Absorption effects of the Earth are considered as part of the detector and folded in the effected area.

Figure 5 provides an overview of effective areas for various analyses that are presented at this conference. First we note that the effective area increases strongly in the range from 100 GeV to about 100

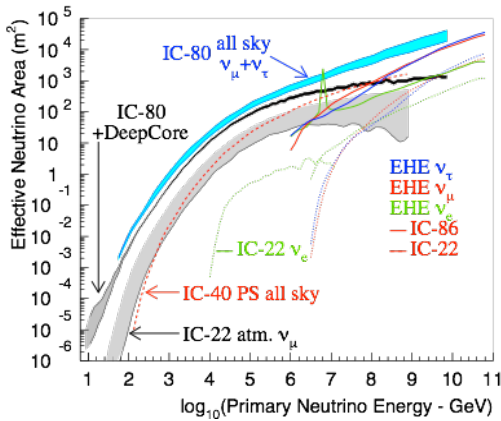

Fig. 5 The neutrino effective area is shown for a several IceCube configurations (IC22, IC 40, IC86), neutrino flavors, energy ranges and analysis levels (trigger, final analysis).

TeV. This is due to the increase in the neutrino-nucleon cross-section and, in case of the muons, the workhorse of high energy neutrino astronomy, due to the additional increase of the muon range. Above about 100 GeV, the increase slows down because of radiative energy losses of muons.

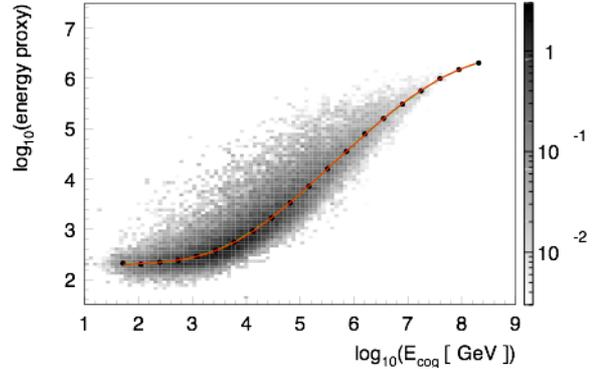

Fig. 6 The energy resolution for muons is approximately 0.3 in log(energy) over a wide energy range

The IC22[8] and IC80 atmospheric $\nu_\mu$ area are shown for upgoing neutrinos. The shaded area (IC22) indicates the range from before to after quality cuts. The effective area of IC40 point source analysis[7] is shown for all zenith angles. It combines the upward neutrino sky (predominantly energies < 1PeV) with downgoing neutrinos (predominantly >1 PeV). Also shown is the all sky $\nu_\mu+\nu_\tau$ area of IC80.

The $\nu_e$ effective area is shown for the current IC22 contained cascade analysis[9] as well as the IC22 extremely high energy (EHE) analysis[10]. It is interesting to see how two entirely different analysis techniques match up nicely at the energy transition of about 5 PeV.

The cascade areas are about a factor of 20 smaller than the $\nu_\mu$ areas, primarily because the muon range allows the detection of neutrino interactions far outside the detector, increasing the effective detector volume by a large factor. However, the excellent energy resolution of contained cascades will benefit the background rejection of any diffuse analysis, and makes cascades a competitive detection channel in the detector where the volume grows faster than the area with the growing number of strings.

The figure illustrates why IceCube, and other large water/ice neutrino telescopes for that matter, can do physics over such a wide energy range. Unlike typical air shower cosmic ray or gamma ray detectors, the effective area increases by about 8 orders of magnitude ($10^{-4} m^2$ to $10^{+4} m^2$) over an energy range of equal change of scale (10 GeV to 10⁹ GeV). The analysis at the vastly different energy scales requires very different approaches, which are presented in numerous talks in the parallel sessions[11, 12, 13, 14, 15, 16, 17, 18, 19, 20, 21, 22, 23, 24, 25, 26, 27, 28, 29, 30, 31, 32, 33, 34, 35, 36, 37, 38, 39, 40, 41, 42, 43, 44, 45].

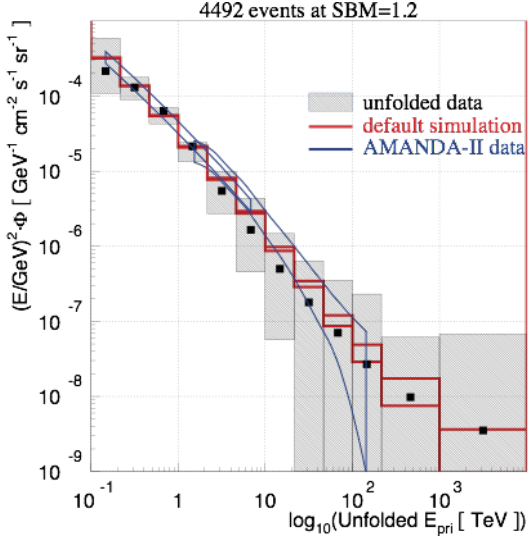

Fig. 7 Unfolded muon neutrino spectrum[8] averaged over zenith angle, is compared to simulation and to the AMANDA result. Data from 22 strings.

The measurement of atmospheric neutrino flux requires a good understanding of the energy response. The energy resolution for muon neutrinos in the IC22 configuration is shown in Fig. 6 [8]. Over a wide energy range (1 – 10000 TeV) the energy resolution is ~0.3 in log(energy). This resolution is largely dominated by the fluctuations of the muon energy loss over the path length of 1 km or less.

## IV. ATMOSPHERIC NEUTRINOS AND THE SEARCH FOR ASTROPHYSICAL NEUTRINOS

We have discussed the effective areas, as well as the angular and energy resolution of the detector. Armed with these ingredients we can discuss some highlights of neutrino measurements and astrophysical neutrino searches.

Figure 7 shows a preliminary measurement obtained with the IC22 configuration. An unfolding procedure has been applied to extract this neutrino flux. Also shown is the atmospheric neutrino flux as published previously based on 7 years of AMANDA-II data. The gray shaded area indicates the range of results obtained when applying the procedure to events occurred primarily in the top or bottom of the detector. The collaboration is devoting significant efforts to understand and reduce systematic uncertainties as the statistics increases. The data sample consists of 4492 high quality events with an estimated purity of well above 95%. Several atmospheric neutrino events are observed above 100 TeV, pushing the diffuse astrophysical neutrino search gradually towards the PeV energy region and higher sensitivity. A look at the neutrino effective areas in Fig. 5 shows that the full IceCube with 86 strings will detect about one order of magnitude more events: ~50000 neutrinos/year.

The search for astrophysical neutrinos is summarized in Fig. 8. While the figure focuses on diffuse fluxes, it is clear that some of these diffuse fluxes may be detected as point sources. Some examples of astrophysical flux models that are shown include AGN Blazars[46], BL Lacs[47], Pre-cursor GRB models and Waxman Bahcall bound[48] and Cosmogenic neutrinos[49].

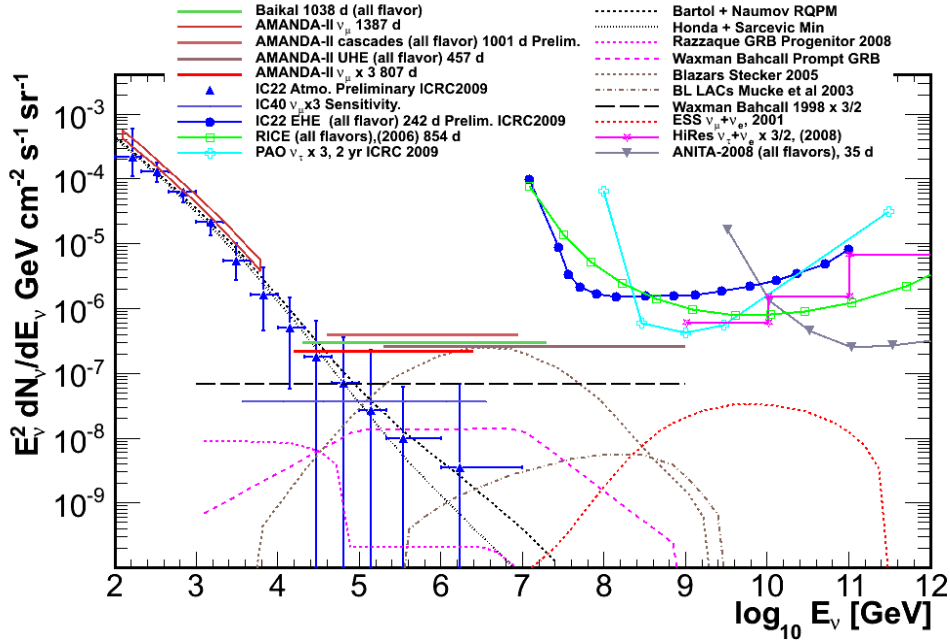

Fig. 8 Measured neutrino atmospheric neutrino fluxes from AMANDA and IceCube are shown together with a number of models for astrophysical neutrinos and several limits by IceCube and other experiments

The following limits are shown for AMANDA and IceCube:

- AMANDA-II, 2000-2006, atmospheric muon neutrino flux[50]
- IceCube-22 string, atmospheric neutrinos, (preliminary)[8]
- AMANDA-II, 2000-2003, diffuse $E^{-2}$ muon neutrino flux limit[51]
- AMANDA-II, 2000-2002, all flavors, not contained events, PeV to EeV, $E^{-2}$ flux limit[52]
- AMANDA-II, 2000-2004, cascades, contained events, $E^{-2}$ flux limit[53]
- IceCube-40, muon neutrinos, throughgoing events, preliminary sensitivity[29]
- IceCube-22, all flavor, throughgoing, downgoing, extremely high energies (10 PeV to EeV)[10]

Also shown are a few experimental limits from other experiments, including Lake Baikal (ref. Avrorin) (diffuse, not contained), and at higher energies some differential limits by RICE, Auger and at yet higher energies energies from ANITA.

The skymap in Fig. 9 shows the probability for a point source of high-energy neutrinos. The map was obtained from 6 months of data taken with the 40 string configuration of IceCube. This is the first result obtained with half of IceCube instrumented. The "hottest spot" in the map represents an excess of 7 events, an excursion from the atmospheric background with a probability of $10^{-4.4}$. After taking into account trial factors, the probability for this event to happen anywhere in the sky map is not significant. The background consists of 6796 neutrinos in the Northern hemisphere and 10,981 down-going muons rejected to the $10^{-5}$ level in the Southern hemisphere. The energy threshold for the Southern hemisphere increases with increasing elevation to reject the cosmic ray the muon background by up to a factor of

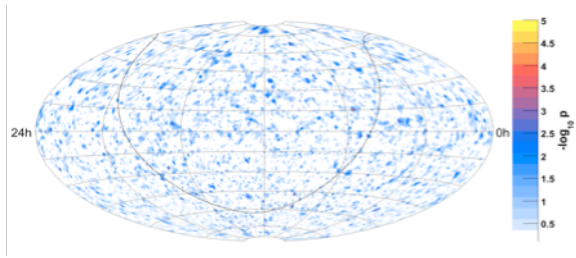

Fig, 9: The map shows the probability for a point source of high-energy neutrinos on the atmospheric neutrino background. The map was obtained by operating IceCube with 40 strings for half a year [6]. The "hottest spot" in the map represents an excess of 7 events. After taking into account trial factors, the probability for this event to happen anywhere in the sky map is not significant. The background consists of 6796 neutrinos in the Northern hemisphere and 10,981 down-going muons rejected to the $10^{-5}$ level in the Southern hemisphere.

$\sim 10^{-5}$. The energy of accepted downgoing muons is typically above 100 TeV.

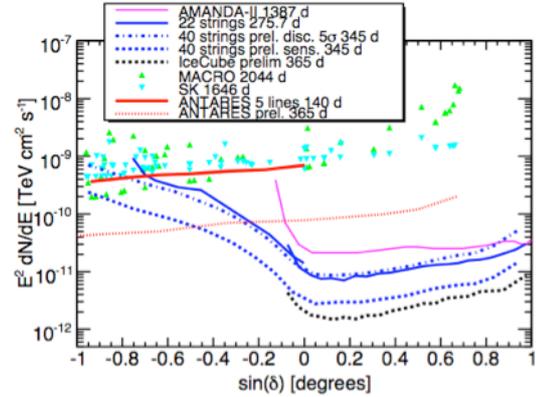

Fig. 10 Upper limits to $E^{-2}$-type astrophysical muon neutrino spectra are shown for the newest result of ½ year of IC40 and a number of earlier results obtained by IceCube and other experiments.

This unbinned analysis takes the angular resolution and energy information on an event-by-event basis into account in the significance calculation. The obtained sensitivity and discovery potential is shown for all zenith angles in the figure.

## V.  SEARCH FOR DARK MATTER

IceCube performs also searches for neutrinos produced by the annihilation of dark matter particles gravitationally trapped at the center of the Sun and the Earth. In searching for generic weakly interacting massive dark matter particles (WIMPs) with spin-independent interactions with ordinary matter, IceCube is only competitive with direct detection experiments if the WIMP mass is sufficiently large. On the other hand, for WIMPs with mostly spin-dependent interactions, IceCube has improved on the previous best limits obtained by the SuperK experiment using the same method. It improves on the best limits from direct detection experiments by two orders of magnitude. The IceCube limit as well as a limit obtained with 7 years of AMANDA are shown in the figure. It rules out supersymmetric WIMP models not excluded by other experiments. The installation of the Deep Core of 6 strings as shown in Fig. 1 will greatly enhance the sensitivity of IceCube for dark matter. The projected sensitivity in the range from 50 GeV to TeV energies is shown in Fig. 11. The Deep Core is an integral part of IceCube and relies on the more closely spaced nearby strings for the detection of low energy events as well as on a highly efficient veto capability against cosmic ray muon backgrounds using the surrounding IceCube strings.

## VI.  COSMIC RAY MUONS AND HIGH ENERGY COSMIC RAYS

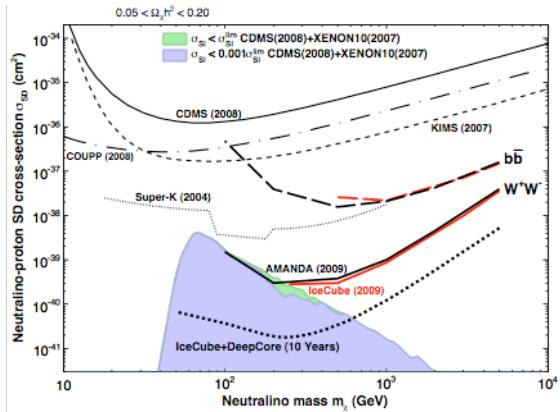

Fig. 11 The red boxes show the upper limits at 90% confidence level on the spin-dependent interaction of dark matter particles with ordinary matter [17. 19]. The two lines represent the extreme cases where the neutrinos originate mostly from heavy quarks (top line) and weak bosons (bottom line) produced in the annihilation of the dark matter particles. Also shown is the reach of the complete IceCube and its DeepCore extension after 5 years of observation of the sun. The shaded area represents supersymmetric models not disfavored by direct searches for dark matter. Also shown are previous limits from direct experiments and from the Superkamiokande experiment

IceCube is a huge cosmic-ray muon detector and the first sizeable detector covering the Southern hemisphere. We are using samples of several billion downward-going muons to study the enigmatic large and small scale anisotropies recently identified in the cosmic ray spectrum by Northern detectors, namely the Tibet array[54] and the Milagro array[55]. Fig. 12 shows the relative deviations of up to 0.001 from the average of the Southern muon sky observed with the 22-string array[11]. A total of 4.3 billion events with a median energy of 14 TeV were used. IceCube data shows that these anisotropies persist at energies in excess of 100 TeV ruling out the sun as their origin. Having extended the measurement to the Southern hemisphere should help to decipher the origin of these unanticipated phenomena.

IceCube can detect events with energies ranging from 0.1 TeV to beyond 1 EeV, neutrinos and cosmic ray muons.

The surface detector IceTop consists of ice Cherenkov tank pairs. Each IceTop station is associated with an IceCube string. With a station spacing of 125 m, it is efficient for air showers above energies of 1 PeV. Figure 13 shows an event display of a very high-energy (~EeV) air shower event. Hits are recorded in all surface detector stations and a large number of DOMs in the deep ice. Based on a preliminary analysis some 2000 high-energy muons would have reached the deep detector in this event if the primary was a proton and more if it was a nucleus. With 1 $km^2$ surface area, IceTop will acquire a sufficient number of events in coincidence with the in-ice detector to allow for cosmic ray measurements up to 1 EeV. The directional and calorimetric measurement of the high energy muon component with the in-ice detector and the simultaneous measurement of the electromagnetic particles at the surface with IceTop will enable the investigation of the energy spectrum and the mass composition of cosmic rays.

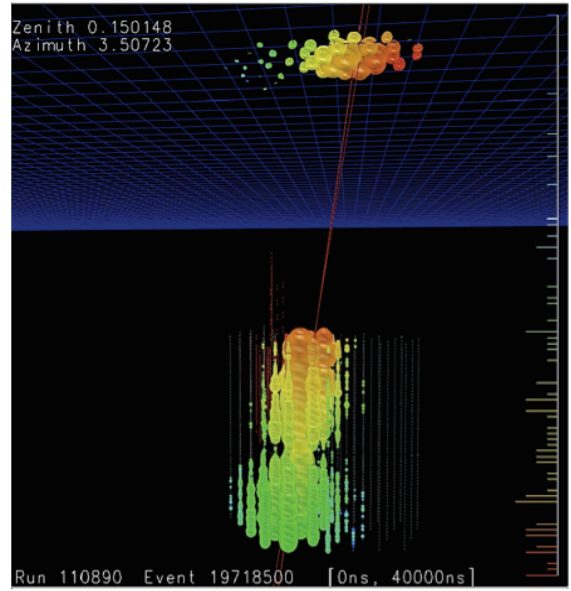

Fig. 13 A very high energy cosmic ray air shower observed both with the surface detector IceTop and the in-ice detector string array.

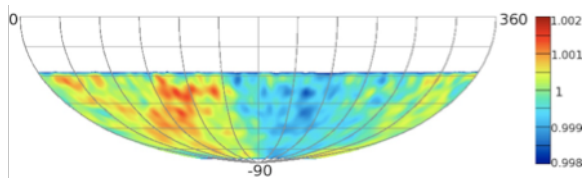

Fig.12 he plot shows the skymap of the relative intensity in the arrival directions of 4.3 billion muons produced by cosmic ray interactions with the atmosphere with a median energy of 14 TeV; these events were reconstructed with an average angular resolution of 3 degrees. The skymap is displayed in equatorial coordinates.

Events with energies above one PeV can deposit an enormous amount of light in the detector. Figure 14 shows an event that was generated by flasher pulse produced by an array of 12 UV LEDs that are mounted on every IceCube sensor. The event produces an amount of light that is comparable with that of an electron cascade on the order of 1 PeV. Photons were recorded on strings at distances up to 600 m from the flasher. The events are somewhat brighter than previously expected because the deep ice below a depth of 2100m is exceptionally clear.

The scattering length is substantially larger than in average ice at the depth of AMANDA.

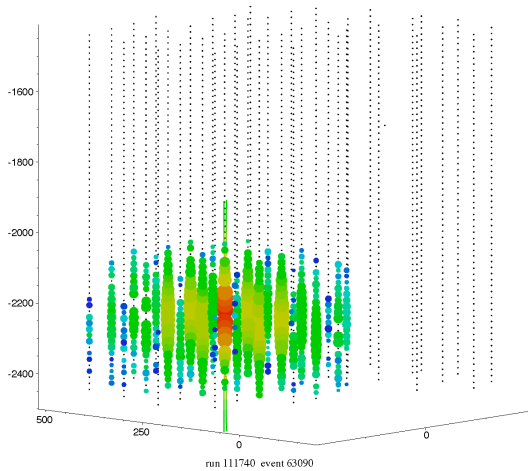

Figure 14: A flasher event in IceCube. Such events, produced by LEDs built in the DOMs, can be used for calibration purposes.

Extremely high energy (EHE) events, above about 1 PeV, are observed near and above the horizon. At these energies, the Earth becomes opaque to neutrinos and one needs to change the search strategy. In an optimized analysis, the neutrino effective area reaches about 4000m$^2$ for IC80 at 1 EeV. IC80 can therefore test optimistic models of the cosmogenic neutrino flux. IceCube is already accumulating an exposure with the current data that makes detection of a cosmogenic neutrino event possible.

## VII. ACKNOWLEDGEMENTS

We acknowledge the support from the following agencies: U.S. National Science Foundation-Office of Polar Program, U.S. National Science Foundation-Physics Division, University of Wisconsin Alumni Research Foundation, U.S. Department of Energy, and National Energy Research Scientific Computing Center, the Louisiana Optical Network Initiative (LONI) grid computing resources; Swedish Research Council, Swedish Polar Research Secretariat, and Knut and Alice Wallenberg Foundation, Sweden; German Ministry for Education and Research (BMBF), Deutsche Forschungsgemeinschaft (DFG), Germany; Fund for Scientific Research (FNRS-FWO), Flanders Institute to encourage scientific and technological research in industry (IWT), Belgian Federal Science Policy Office (Belspo); the Netherlands Organisation for Scientific Research (NWO); M. Ribordy acknowledges the support of the SNF (Switzerland); A. Kappes and A. Groß acknowledge support by the EU Marie Curie OIF Program.